# Mixing enhancement induced by viscoelastic micromotors in microfluidic platform

*A. Zizzari[1,\*], M. Cesaria[2,3,\*], M. Bianco[1], L. L. del Mercato[1], M. Carraro[4], M. Bonchio[4], R. Rella[3], V. Arima[1]*

[1]CNR NANOTEC - Institute of Nanotechnology, c/o Campus Ecotekne, Via Monteroni, 73100 Lecce, Italy

[2]University of Salento, Department of Mathematics and Physics "E. De Giorgi" via Arnesano, 73100 Lecce, Italy

[3]CNR IMM - Institute for Microelectronics and Microsystems, c/o Campus Ecotekne, via Monteroni, 73100 Lecce, Italy.

[4]ITM-CNR and Department of Chemical Sciences, University of Padova Via Marzolo 1, 35131 Padova, Italy.

*Corresponding authors: alessandra.zizzari@unisalento.it; maura.cesaria@le.infn.it.

**ABSTRACT**

Fine manipulation of fluid flows at the microscale has a tremendous impact on mass transport phenomena of chemical and biological processes inside microfluidic platforms. Fluid mixing in the laminar flow regime at low Reynolds is poorly effective due to the inherently slow diffusive mechanism. As a strategy to enhance mixing and prompt mass transport, here, we focus on polyelectrolyte multilayer capsules (PMCs) embodying a catalytic polyoxometalate as microobjects to create elastic turbulence and as micromotors to generate chaotic flows by fuel-fed propulsions.

The effects of the elastic turbolence and of the artificial propulsion on some basic flow parameters, such as pressure and volumetric flow rate are studied by a microfluidic set-up including pressure and flow sensors. Numerical-handling and physical models of the experimental data are presented and discussed to explain the measured dependence of the pressure drop on the flow rate in presence of the PMCs. As a practical outcome of the study, a strong decrease of the mixing time in a serpentine microreactor is demonstrated.

Unlike our previous reports dealing with capillarity flow studies, the present paper relies on hydrodynamic pumping experiments, that allows us to both develop a theoretical model for the understanding of the involved phenomena and demonstrate a successfully microfluidic mixing application.

All of this is relevant in the perspective of developing microobject based methods to overcome microscale processes purely dominated by diffusion with potential improvements of mass trasport in microfluidic platforms.

***Keywords:*** *Microfluidics, mass transport, mixing, micromotors, polyelectrolyte capsules, polyoxometalate*

**1. Introduction**

Integrated microfluidic platforms for lab-on-a-chip (LOC) and organ-on-a-chip (OOC) applications are of fast growing applicative interest with huge potential for the development of laboratory scale and industrial products. Compared to bulky and traditional platforms, LOCs and OOCs possess many potentials and benefits (small consumption of reagents and samples, short reaction times, highly parallel and multiplexed analysis and portability) which make them useful for biochemical analyses, chemical processing applications [1, 2], environment and sanitation safety [3], drug delivery, cell culture manipulation [4-6]. Nowadays, due to the important advances in fabrication technologies and miniaturization, LOCs and OOCs can be realized with sub-micrometer level of precision and reproducibility. Moreover, significant improvements in terms of device performance are due to the microfluidic technologies that support a precise control of small quantities of liquids flowing through LOCs and OOCs. However, some aspects of fluid control at the microscale limit further advancing in the applications of microfluidic platforms. Indeed, high surface-to-volume ratios and significant surface tension can cause low diffusivity, inefficient thermal conduction and a great increase of hydrodynamic drag [7]. Furthermore, at the microscale laminar flow with low Reynolds [8-10] and low Péclet numbers dominates, indicating that the mass transport by means of fluid displacement is essentially driven by diffusion and standard microfluidic flows behave similarly to flows of highly viscous fluids [11]. Consequently, fluid mixing operation based only on the inherently slow diffusive mechanism requires a long channel for effective mixing [12]. This issue is a serious drawback for many microfluidic applications, where an excellent mixing of liquids is critical to be accomplished over short distances (a few centimeters) [13, 14]. For example, in a biomedical assay or an *in vitro* model mimicking a physiological environment, rapid mixing between fluids containing reagents or nutrients is necessary to achieve a fast diagnosis or to guarantee cell survival. Thus, in order to facilitate mass transport at the microscale, a number of different working technologies have been developed, such as numerous designs of *active* and *passive* micromixers with the aim of increasing Reynolds numbers towards turbulent regimes [15-21] or reduce the species diffusion length in a laminar regime by increasing the interfacial contact area between different flows [22]. Although the passive mixers can be easily incorporated into a chip during its fabrication, users cannot easily manipulate the level of mixing enhancement. On the other hand, even if active mixing enhancement offers superior control over the level of mixing, more complex and expensive fabrication methods and external power sources are needed [23].

Alternatively, viscoelastic polymers [24-26] or microvesicles [27, 28] are reported to generate elastic turbulence resulting in non-linear dynamics of flow with a prospective application to improve the fluid mixing. Air bubbles excited by a single ultrasonic wave were also proposed as microfluidics mixer, given their attitude to generate collective flows [29]. Based on the same effect, the use of a microswimmer was suggested to perform biosensing, drug delivery, imaging, and microsurgery in *in vitro* microfluidic environments or *in vivo* human bodies narrow passages [30].

Recently, self-propelled autonomous nano/micromotors, in particular structures with high speed and good biocompatibility, have attracted a so great interest in the field of material science and technology research that their applications are gradually expanding [31, 32]. Some researchers, inspired by biological enzymatic processes, introduced self-propelled catalytic nano/micromotors, as functional actuators in microfluidic channels [33]. Furthermore, in a theoretical study on micron-scale rigid beads confined by two walls of a microchannel in a low Reynolds number fluid, micromotors have been proposed as active micromixers [34].

Accounting for the wide interest in and promising applications of micromotors, herein we experimentally investigate and discuss on a fundamental basis the use of micromotors for mixing purposes in a microfluidic platform. As model system, polyelectrolyte multilayer capsules (PMCs) integrating a Ru-substituted polyoxometalate ($Ru_4POM$) were investigated. $Ru_4POM$ and hydrogen peroxide ($H_2O_2$) react accordingly to a bimolecular reaction, to yield $H_2O_2$ dismutation with production of oxygen and water [35]:

$$2\ H_2O_2 \rightarrow O_2 + 2\ H_2O$$

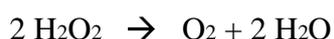

In the surrounding carrier fluid, this results in microcapsule self-propulsion and gas bubbles production. Owing to the high stability of the catalyst (catalytic turnover number, TON> $3 \times 10^5$), the reaction proceeds up to complete $H_2O_2$ consumption in neutral water. Considering that no residual catalytic oxidant activity, resulting from $H_2O_2$ activation, was observed, the method is convenient for aqueous solutions containing $H_2O_2$-stable substrates and analytes.

Compared to the rigid inorganic micromotors discussed in ref. [33], soft organic micromotors are expected to increase mass transport by combining manifold effects: 1) considering their intrinsic viscoelasticity, they can generate elastic turbulence; 2) by activation through suitable stimulus, they are able to move in the fluid like swimming micro-organisms and produce gas bubbles; 3) gas bubbles motion can be further enhanced by applying ultrasonic waves.

In our previous works, the $H_2O_2$ dismutation by means of $Ru_4POM$-PMCs was studied in a capillary flow regime with a focus on kinematics quantities (velocity and acceleration of the fluid) [36, 37] and the pressure-driven turbulence effects (corresponding to an applied pressure of tens of Pa) were demonstrated to blow up flexible thin membranes for gas sensing [38] or liquids pumping applications [39]. As further step, stimulated by the considerable development of applications of micromotors in the pharmaceutical, biomedical and chemical fields, in this work we evaluated the impact of activated PMCs on the fluid mixing under hydrodynamic pumping conditions, which, unlike capillary flow regime, are operative common working flow conditions. Compared to viscoelastic polymers, that are miscible with the working liquid and hence difficult to remove from the mixture, soft PMCs can be easily removed from the reaction environment via centrifugation or external magnetic fields whenever decorated with magnetic nanoparticles. Moreover, when stimulated by a proper chemical, PMCs act as self-propelled micromotors due to an oxygen bubble-driven mechanism where the propulsion is due to pressure changes and recoil effects favored by bubble generation in the presence of $H_2O_2$ fuel source. Hence, capsules moving in the background fluid speed up the diffusion-limited mass transport through detaching or bursting bubbles from their surface.

An experimental study of the effect of $Ru_4POM$-PMCs on fluid flows is presented both in the absence and in the presence of the $H_2O_2$ (the stimulus activating the self-propellant effect of the micromotor) and depending on the position (inner or external layer) of the $Ru_4POM$ material loaded in PMCs. The dependence of the pressure drop on the flow rate was correlated to the PMCs viscoelastic and micromotor properties thanks to a theoretical model based on the relationship between pressure drop and capillary number depending on the flow rate values. We demonstrated that the overall effect produces an effective enhancement of fluid mixing, which is particularly promising at the applicative level for LOC and OOC platforms to improve mass transport in experimental conditions purely dominated by diffusion.

## 2. Materials and methods

### 2.1 Materials

For the fabrication of microchannels, commercially available B-270 glasses, cut in 2.5 cm x 2.5 cm x 1.1 mm (width x depth x height) slides and covered with 450 nm-thick Chromium (Cr) layer (Telic, USA), were used as solid substrates. Hydrochloric acid (HCl), ammonium fluoride ($NH_4F$), and hydrofluoric acid (HF) were purchased from Sigma–Aldrich (Taufkirchen, Germany). The resist AZ9260 and the AZ400k developer were purchased from MicroChemicals (Ulm, Germany). The chromium etchant solution was purchased from Poletto Aldo s.r.l. (Noventa di Piave, Venice, Italy). Photomasks were designed using CleWin software, and printed by J. D. Photo-tools Ltd. (Oldham, Lancashire, UK). Fluoropolymer tubing (Tub FEP Blu 1/32 x 0.09) were purchased from IDEX Health & Science (Germany).

For the preparation of PMCs, milli-Q water ($H_2O$) with a resistivity of 18.2 MΩ cm was used. Poly(sodium 4-styrenesulfonate) (PSS, Mw ≈ 70 KDa), poly(allylamine hydrochloride) (PAH, Mw ≈56 KDa), calcium chloride dehydrate ($CaCl_2$, Mw = 147.01 Da), sodium carbonate ($Na_2CO_3$, Mw = 105.99 Da), ethylenediaminetetraacetic acid disodium salt dehydrate (EDTA) were purchased from Sigma–Aldrich (Milan, Italy).

For the mixing experiments, hydrogen peroxide ($H_2O_2$, 30%) and eosin were purchased from Sigma-Aldrich (Milan, Italy).

### 2.2 Preparation of catalytic PMCs

The synthesis of $Na_{10}[Ru_4O_4(OH)_2(H_2O)_4(\gamma-SiW_{10}O_{36})_2]$ (Ru4POM) was performed following a procedure reported in detail elsewhere [40]. Two reactive batches of PMCs were prepared by adsorbing the anionic catalytic $Ru_4POM$ either as inner layer (batch termed B1, Figure 1a) or as external layer (batch termed B2, Figure 1b) during layer-by-layer assembly procedure [41-43]. Briefly, ≈20 mg of calcium carbonate ($CaCO_3$) microparticles were sequentially resuspended for 12 min in 1 mL of 0.5 M NaCl solution containing the polyanion PSS (2 mg $mL^{-1}$, pH = 6.5) or the polycation PAH (2 mg $mL^{-1}$, pH = 6.5). The excess polyelectrolytes were removed by three washing with 1 mL of milli-Q $H_2O$ and three centrifugation steps (4500 rpm for 5 s). For preparing PMCs of batch B1, the microparticles coated with two bi-layers of (PSS/PAH) were incubated in 1 mL of $Ru_4POM$ solution ($H_2O$, pH = 5.0, 2 mg $mL^{-1}$, 0.35 mM). Then, three additional bi-layers of (PSS/PAH) were deposited to obtain the following multilayer shell composition (PSS/PAH)($Ru_4POM$/PAH)$_2$(PSS/PAH)$_3$ (batch B1, Figure 1a). Instead, for obtaining PMCs of batch B2, the microparticles carrying five bi-layers of (PSS/PAH) were incubated in 1 mL of $Ru_4POM$ solution ($H_2O$, pH = 5.0, 2 mg $mL^{-1}$, 0.35 mM) to obtain the following multilayer shell (PSS/PAH)$_5$($Ru_4POM$) (batch B2, Figure 1b). PMCs without $Ru_4POM$ (batch termed B3) were also prepared and used as control system (Figure 1c). The sacrificial $CaCO_3$ cores were removed by complexation with EDTA buffer. Finally, the PMCs were stored as suspension in 500 μL of milli-Q $H_2O$ at 4 °C. All obtained PMCs showed a spherical shape, with average diameter of 4.00 ± 0.04 μm (Figure S1a-c). The PMCs number per volume was determined by direct counting by a haemocytometer under a microscope by using a 20X objective in phase contrast channel. The number of PMCs was estimated to be around $9.05 \times 10^8$ per mL in batch B1, $1.12 \times 10^9$ PMCs per mL in batch B2, and $7.8 \times 10^8$ PMCs per mL in batch B3. The uptake of the $Ru_4POM$ in batches B1 and B2 was estimated by UV-Vis analysis (Cary 300 UV-Vis, Varian). It was calculated from the absorbance difference (λ=270 nm, $\varepsilon_{POM}$ = 74846 $M^{-1} cm^{-1}$) between initial feeding solution (total $Ru_4POM$) and the supernatants recovered after the incubation step of microparticles with the feeding solution of $Ru_4POM$ (free $Ru_4POM$) (Figure S1d). A total amount of 203 μM and 189 μM of $Ru_4POM$ was estimated for batch B1 ($9.05 \times 10^8$ per mL) and for batch B2 ($1.12 \times 10^9$ PMCs per mL).

## 2.3 Experimental methods

### 2.3.1 Flow sensor –based measurements in a linear microchannel

In perspective of mass transport in pure laminar regime, we analyzed the effects on flow dynamics of all the three PMC batches (B1, B2 and B3) inside a glass linear microchannel under hydrodynamic pumping conditions. In particular, we monitored pressure drop and volumetric flow rate based on a set-up including microfluidic flow sensors (Elveflow®, ELVESYS S.A.S, France, see Figure S3) connected to a 1.5 cm long glass microchannel patterned with a nominal width of 300 μm by using photolithography on a B-270 glass substrate [44]. After the geometry transfer, the glass substrate was etched with buffered oxide etch (BOE) solution by using the microwave reactor system (Anton Paar Multiwave 3000, Labservice Analytica s.r.l., Italy) as reported elsewhere [45] for obtaining a channel depth of 60 μm. Then, two holes were processed by using microdriller (MICRO miller MF70, Proxxon, Germany), in order to create inlet/outlet ports to be used as connections to the Elveflow® sensors. The channel was then thermally bonded to a glass top plate [46, 47]. Based on the same fabrication procedure of the linear glass channel, a 270 mm long, 400 µm wide and 100 µm deep serpentine glass microreactor was prepared for mixing experiments.

For each experiment, an aqueous solution containing 5M $H_2O_2$ and 1.9 x $10^3$ cps $\mu L^{-1}$ (with ca 0.43 and 0.32 μM $Ru_4POM$ adsorbed in the case of B1 and B2) was prepared. The $Ru_4POM$ act as a $H_2O_2$ dismutation catalytic trigger, thus providing localized oxygen gas bursts to generate turbulence [34]. The solution was allowed to flow in the network of the assembled Elveflow® sensors-microchannel set-up at different values of the inlet volumetric flow rate ($Q_{in}$ = 10, 15, 20, 30, 40, 50, 60 μL $min^{-1}$). The comparison between reactive and not reactive solutions was performed at constant number of capsules to allow the evaluation of the contributions due to the physical properties of capsules and to the reaction.

The concentration and the volumetric flow rate of the solution were chosen consistently with the characteristics of our sensors, that is by considering their pressure and flow speed ranges (0-8 bar and 2 -80 µL/min respectively).

Under such dynamic conditions, the inlet and outlet sensors monitored the inlet pressure ($P_{in}$), the outlet pressure ($P_{out}$) and the volumetric output flow rate ($Q_{out}$) during a designed time of two minutes (long enough to observe multiple events of peaked pressure). The evolution of the inlet volumetric flow rate was also monitored to assess its temporal stability and consistency with the nominal value set in our experiments. The sensors were characterized by sensitivity in the range 2 – 60 µL min$^{-1}$. The pressure drop along the microchannel, as well as $Q_{out}$ of the reactive batches (B1, B2), the non-reactive batch (B3) and the transport fluid ($H_2O$), were compared and analyzed using OriginPro 8.0 software to handle and plot the measured data.

**2.3.2 Microfluidic experiments for mixing applications**

The two reactive mixtures (containing B1, B2) and the non-reactive one (with B3) were also used to test the PMCs effects on mass transport during diffusive mixing of fluids in a serpentine glass microreactor (270 mm long and with internal volume of 4 µL). Solutions of PMCs (B1, B2 and B3) and $H_2O_2$, prepared as above reported, were red colored using eosin for an optimal visualization of the laminar flow and diffusive mixing process. Each sample was injected into one inlet port of the serpentine microreactor at a flow rate of 50 µL min$^{-1}$ while pure water has been flowing at the same rate through the other inlet port.

The performance of each PMC solution, in terms of mixing efficiency, was evaluated by following the flows along the serpentine channel with an optical microscope (NIKON mod. DS-5MC camera). The mixing length was estimated by monitoring the color change of an eosin solution in the microreactor in comparison with the one observed when eosin/$H_2O_2$ without PMCs and aqueous solutions flow. The images were acquired at a residence time ($R_t$) 0 s, 0.5 s and 3.24 s for each solution of PMCs and compared with images acquired in absence of PMCs. These images were converted to grayscale images and analyzed by using ImageJ software to accurately estimate the efficiency and length-scale of the mixing. In particular, for each image the average change in intensity profile was examined over several lines perpendicular to the channel axis drawn along the channel. All the average values were normalized to the intensity of the colored stream before mixing and plotted *versus* off-center distances (representing the distance off the longitudinal centerline of the microchannel) [21]. The data reproducibility was checked in different chips. Moreover, a coefficient of variation (CoV) ≡ $\sigma/\mu$ [21], defined as the ratio of the standard deviation of the concentration to the mean value of the various concentrations at each datum point, was determined. This ratio, related to the mixing efficiency, was calculated at the inlet region, first meander and middle of the serpentine. It is exactly zero for complete mixing and 1 in absence of mixing. We calculated CoV as average and standard deviation of the values obtained from three experiments performed on three different devices for each mixture of PMCs and for the solution without PMCs used as reference.

## 3. Results and discussion

### 3.1 Experimental data analysis

In order to rationalize our experiments and their aim we remark a few procedure points.

Since the motion inertia typical of the capillary-flow regime stops the flow inside the microchannels over short distances from the inlet section, such working conditions are interesting for a limited number of LOCs [37] in which an efficient mixing of reagents is not needed. Hence, differently from our kinematic studies under capillary-flow regime [35,36], in this work we analyze the effects of the catalytic activity of PMCs inside a glass linear microchannel under hydrodynamic pumping conditions. This issue is of most applicative interest for LOCs and OOCs relying on mass transport processes in continuous flow.

Moreover, in addition to the PMCs loaded with the Ru$_4$POM as inner layer (B1), we also explored the behaviour of the capsules with the catalyst adsorbed on the external layer (B2). This comparison would let evaluate the influence of the spatial confinement of the catalyst between the polyelectrolyte multilayers on the mixing efficiency.

To investigate physics underlying our experimental findings, our results are presented and discussed according to the following guidelines and interpretative analysis steps.

i) As a preliminary investigation, the behavior of the flowing pure water (i.e., the carrier fluid of PMCs) as control system was modeled to investigate the hydrodynamics of milli-Q water (H$_2$O) flowing inside the microchannel in laminar regime steady-state conditions (Figure S2 in SI). The results were compared with the corresponding measured quantities to confirm the quality and well-working of the etched linear microchannel (Figure S2 and S4 in SI and corresponding discussion).

ii) The effect of non-reactive capsules (i.e., B3) in presence of H$_2$O$_2$ on the carrier fluid (H$_2$O) was tested by defining the pressure drop of B3 normalized to the pressure drop of the transport fluid, that is

$$\Delta P\ (B3, H_2O) = \frac{\Delta P\ (B3) - \Delta P\ (H_2O)}{\Delta P\ (H_2O)} \qquad (1)$$

where ΔP stands for the pressure difference between inlet and outlet pressure values measured by the microfluidic sensors connected with the inlet and outlet of the microchannel (experimental set-up in Figure S3 in SI).

(iii) The effect of the dismutation reaction resulting from vigorous O$_2$ production was tested by defining the pressure drop of the reactive samples (i.e., B1 and B2) normalized to the above defined pressure drop of B3, that is

$$\Delta P\left(\frac{B1,2}{B3}\right) = \frac{\Delta P\ (B1,2) - \Delta P\ (B3)}{\Delta P\ (B3)} \qquad (2)$$

In regard to the normalization protocol reported in equation 1, it is worth noticing that it is not a limit and doesn't affect the predictions and results of our study.

Indeed, as the inset of Figure 2(a) shows, the experimental pressure-drop curves of the not reactive batch B3 and pure water are nearly overlapping over the whole sampled range of $Q_{in}$, with slight differences/fluctuations for increasing flow rate due to capsule-induced viscoelastic effects (as it will be detailed in Paragraph 3.2). Hence, our experiments indicate that the pressure drops associated with $H_2O$ and $H_2O_2$ are experimentally comparable/undistinguishable. As a confirmation, viscoelastic perturbation induced by PMCs in pure water have been observed in absence of $H_2O_2$ (see Figure 2C plot of ref. [37]). Additionally, microbubbles dispersed in aqueous solutions [54] exhibits a pressure drop consistent with our experiments involving B3 in $H_2O_2$.

Therefore, in practice, there is no difference in using $H_2O$ rather than $H_2O_2$ as control fluid for measuring the pressure drop versus $Q_{in}$ of unreactive capsules through formula (1). Conceptually, since i) capsules rather than $H_2O_2$ determine the behavior of B3, ii) the concentration of $H_2O_2$ changes over time in the case of B1 and B2, and iii) $H_2O$ is the common steady background of all batches under study, we preferred to introduce normalization with respect to $H_2O$ rather than to the fuel.

**3.2 Pressure drop distribution: interpretative model of the experimental data**

The microfluidic sensors connected with the inlet and outlet of the microchannel allowed us to measure the temporal evolution of inlet and outlet pressure ($P_{in}$ and $P_{out}$) and flow rate ($Q_{in}$ and $Q_{out}$) during the experiments involving pure water, non-reactive capsules and the dismutation reaction driven by $H_2O_2$.

In regard of the carrier fluid, it is worth noticing that the observed consistency between the modeled and experimental behavior of our channel under water flow (see ESI) involves that the results associated with the channel in presence of B1, B2 and B3 are expected to depend on the capsules and oxygen-related propulsion phenomena.

The trend of some representative curves of the temporal evolution over 2 minutes of the pressure drop ΔP at fixed $Q_{in}$ are reported in Figure 2a for B1, B2 and B3 during the reaction with $H_2O_2$, and pure water as reference. Several measurement sets were acquired with freshly prepared solutions to check the effectiveness of the dismutation reaction (random occurrence of pressure peaks in the presence of reactive batches) and the general features of the phenomena hereafter discussed. Least square linear fits were drawn for each set of data (red lines), in order to extrapolate a quantitative estimation of $\Delta P$ for each fixed $Q_{in}$.

Unlike water and B3, the catalytic reaction fueled by $H_2O_2$ in the presence of $Ru_4POM$ embodied in the capsules generates oxygen bubble-driven turbulence phenomena and, consequently, an unpredictable fluid behavior that demands a physically consistent method to handle the experimental data.

Turning to the reactive samples (B1 and B2 solutions), $O_2$-bubble generation (see fluorescence images of B1/B2 and B3, prepared as reported in the ref [36], in the inset of Figure 2a) caused temporal fluctuations of the measured quantity leading to a non-linear trend. In this case, to allow a meaningful numerical handling of the measured pressure curves, linear fits were performed in such a way that an average linear trend was individuated as background reference behavior around which the reaction-induced fluctuations distribute. The $\Delta P$ values reported in Figure 2b were all extrapolated based on such procedure. By comparison with pure water, the pressure drop associated with B3 exhibits a slight decrease for $Q_{in} < 15$ µL/min and a monotonous increase for larger inlet flow rate values. This behavior will be discussed in detail on the basis of the interpretative model developed in the paragraph 3.2 as related to the influence of different flow-rate regimes on the motion of the capsules.

It can be clearly observed that B1 (which is associated with inner $Ru_4POM$ layer) leads to more effective changes in $\Delta P$ than B2 (which is associated with external $Ru_4POM$ layer) as a function of the ingoing flow rate. On the other hand, the presence of non-reactive capsules (i.e., B3) provides no oscillating temporal variation in the distribution of $\Delta P$ versus $Q_{in}$ ranging from 10 to 60 µL/min.

The differences observed for B1, B2 and B3 are consistent with what expected based on the different mixing mechanisms active in the non-reactive versus reactive samples. In general, the presence of inert capsules in a carrier fluid causes temporal variation of the pressure distribution with both segmentations of the continuous flow and recirculation zones. Moreover, viscous resistance to flow develops in the channel depending on the flow rate. Whenever capsules are smaller than the cross section of the channel, a thin wetting film of the carrier fluid separates the capsules from the wall of the channel and favors increased viscous dissipation leading to a speed of the capsules different from the one of the background fluid. As a consequence of this disparity, the portion of the continuous fluid that is confined between the capsules can develop convection rolls which favor mixing. Also, the increase of resistance due to the capsules can be exploited to mix the continuous fluid [48].

Differently, in the presence of bubbles generated through reactions between the motor material ($Ru_4POM$) and the solution fueled by $H_2O_2$, mixing is enhanced and prompted by a random recoil effect caused be the continuous growth and ejection of bubbles from the reactive capsules.

This chaotic mechanism is inherently more effective for mixing than capsules stretching and folding the continuous liquid.

Turning back to the $\Delta P$ profile *versus* $Q_{in}$ for B3 as plotted in Figure 3a, corresponding to the green curve of Figure 2b, it clearly shows the occurrence of two different regimes: decreasing $\Delta P$ with a minimum value at 15 µL min$^{-1}$ and then increasing $\Delta P$ with a non-linear trend. The $Q_{out}$ *versus* $Q_{in}$ curve (Figure 3b) has a similar profile but with a minimum at 10 µL min$^{-1}$. This not linear dependence of $Q_{out}$ *versus* $Q_{in}$ is consistent with non-linear increase of the system velocity in presence of PMCs described elsewhere [37].

In order to explain the $\Delta P$ vs $Q_{in}$ curves on a physical basis, the best fitting curve based on the least square method was determined for each region with $Q_{in}$ < 15 µL min$_{-1}$ and $Q_{in}$ > 15 µL min$_{-1}$, as shown in Figures 3c-d.

While for $Q_{in} \leq 15$ μL min$^{-1}$ ΔP decreases matching a parabolic law (see Figure 3c, red continuous line), for $Q_{in} > 15$ μL min$^{-1}$ the ΔP profile can be decomposed in different functional forms depending on the $Q_{in}$ range. In detail, the measured ΔP (referred to as (ΔP)$_{exp}$ in the graph (Figure 3d), similarly to the theoretical pressure drop, follows a linear trend for $15 < Q_{in} \leq 25$ μL min$^{-1}$ (red full circles in Figure 3d) and a power dependence having exponent (5/12) for $Q_{in} > 25$ μL min$^{-1}$ (black full squares in Figure 3d) with better and better accordance for increasing $Q_{in}$.

To explain such behavior that significantly differs from that of water (shown as black curve in Figure 2b), theories about motion of capsules along a pipe have to be taken into consideration. The motion of capsules in a pipe can be classified into different regimes which depend on the flow speed and the capsule density [49]. Based on the reported regimes, once $Q_{in}$ is defined as the bulk fluid velocity, our results suggest that at injection-flow below 15 μL min$^{-1}$, $Q_{in}$ is too low, meaning that an insufficient drag is developed on the capsules to overcome the contact friction between the capsules and the pipe and allows the capsules to slide and move away from the channel inlet. Consequently, capsules rest on the pipe floor and hinder the flow, hence decreasing the $Q_{out}$ and determining a pressure drop ΔP. This behavior is confirmed by the excellent parabolic fit shown in Figure 3c; the quadratic term is physically related to the damping effect due to the drag force. At $15$ μL min$^{-1} \leq Q_{in} \leq 25$ μL min$^{-1}$, another regime is observed that can be ascribed to velocity of the fluid high enough to cause the capsules to slide along the pipe ($Q_{in} > Q_i$, where $Q_i$ is the incipient flow at which the capsule starts to slide). Once the motion of the capsules onsets, depending on the value of $Q_{in}$, it is expected that the liquid bypasses or transports the capsules. If the fluid velocity is still relatively low, the contact friction between the capsules and the pipe is high, and the capsule velocity is less than the fluid velocity as well as the capsules reduce the volumetric flow rate. Consistently, in our experiments, $Q_{out}$ is lower than $Q_{in}$ and ΔP starts to increase linearly since the drag force is negligible. At $Q_{in} > 30$ μL min$^{-1}$, another regime onset. Indeed, for increasing bulk velocity $Q_{in}$, the pressure drop along the capsules (propelling force) would be expected to be higher than the pressure drop felt by the counterpart flow free of capsules [49]. Hence, the capsule velocity should overtake the fluid velocity. All of this is consistent with the experimental results obtained under capillary-driven motion of B1, where speed up of the fluid motion was observed for reactive capsules in absence of $H_2O_2$ [37]. The least square approach leads to a polynomial law with a $Q_{in}^{5/12}$ dependence (Figure 3d). The not linear dependence of $\Delta P$ on the capillary number Ca (and consequently on the flow speed $Q_{in}$) has been already reported in the case of bubbles flowing along a rectangular channel [50] and attributed to the deformability of the bubbles [51-53].

In the case of a fluid transporting bubbles along a rectangular channel, the total pressure drop $\Delta P_{tot}$ can be decomposed as the sum of the pressure drops around each bubble, the body of the bubble and the region of the channel free of bubbles [50, 52, 53]:

$$\Delta P_{tot} = a\, Ca + b\, Ca^{2/3} \qquad (3)$$

where $C_a = \mu V_l \gamma^{-1}$ is the capillary number ($\mu$ and $V_l$ is the viscosity and the velocity of the fluid, respectively, and $\gamma$ is the fluid-bubble interfacial energy), *a* and *b* represents some constant values, and the curvature of the capsule determines the dependence $C_a^{2/3}$. While the linear scaling law is expected to dominate for small values of $C_a$, power-law dependence with exponent 2/3 would occur at high $C_a$. Since $C_a$ is proportional to $Q_{in}$, all of this is fully consistent with the different regimes of pressure drop observed in our experiments at the lowest and highest sampled $Q_{out}$ (Figure 3d).

At low flow rate the curvature of the capsules is not affected by the transport fluid [54]; for increasing flow rate the viscoelastic effects begin to play a role [37] and at high flow rate a plug flow regime occurs where the fluid transports the capsules [54].

In our experiments, the observed decomposition of the non-linear part of $\Delta P$ ($\Delta P_{NL}$) given by

$$\Delta P_{NL} = k\, Q_{in}^{5/12} = f(Q_{in}) Q_{in}^{2/3} \qquad (4)$$

indicates a dependence of the interfacial energy on the fluid velocity (i.e., $f(Q_{in})$ including a constant *k* multiplied by a $Q_{in}$ dependent term) that reflects an intermediate regime influenced by the interactions at the fluid-capsule interface [55]. These interactions are responsible of the $Q_{in}$-dependent modulated amplitude of the dependence $Q_{in}^{2/3}$ measured in our experiments. Consistently, the "soft" material (polyelectrolytes) composing our capsules are expected to favour interactions at the fluid-capsule interface more effective than the ones developing at the interface between a transport-fluid and a capsule deformable for increasing $Q_{in}$.

This picture is consistent with the observed overlapping between the experimental data $(\Delta P)_{exp}$ and the law $Q_{in}^{2/3}$ in Figure 3d for $Q_{in} > 30$ μL min$^{-1}$, as well as with the pressure drop of B3 normalized to the one of the transport fluid (Figure 4a) calculated by the formula (1), where $\Delta P = P_{in}-P_{out}$. Indeed, at the highest sampled $Q_{in}$, constant $\Delta P(B3, H_2O)$ clearly demonstrates the dominance of the transport mechanism of the capsules by the background fluid.

Regarding the non-linear behavior of $Q_{out}$ *versus* $Q_{in}$ shown in the Figure 3b, at very low inlet flow rate, the high contact friction between the capsules and the channel wall causes a slow motion of the capsules that preferentially rest on the channel floor and hinder the flow, hence decreasing $Q_{out}$ significantly. Once capsules are able to move, increasing $Q_{in}$, drives the system in the regime where the capsule velocity is lower than the fluid velocity. Hence, the presence of the capsules causes $Q_{out} < Q_{in}$. For further increasing $Q_{in}$, the non-linear increase of the pressure drop (Fig. 3(d)) related to the viscoelasticity/deformability of the capsules in the background flux, accounts for $Q_{out} < Q_{in}$.

Turning to the reactive solutions (B1 and B2), Figure 4b represents the increase of $\Delta P$ due to the $H_2O_2$–driven dismutation reaction. The pressure values reported in the graph, i.e., $\Delta P(B1,2/B3)$ as defined by formula (2), represent the pressure drop $\Delta P = P_{in}-P_{out}$ of the reactive batches (B1 and B2) normalized with respect to the pressure drop of B3. Based on its definition, $\Delta P(B1,2/B3)$ lets extrapolate the propelling effect due to the $O_2$ production and compare B1 and B2 in terms of the location of the active Ru$_4$POM material. The maximum effect is clearly observed at 15 μL min$^{-1}$ for both the reactive solutions where the effects of the PMCs viscoelasticity are minimum (see Figure 4b). The graph also highlights the superior performance of B1 PMCs, where the slight higher amount of Ru$_4$POM sandwiched within the multilayer shell. Because of the difficulties to fit the original data at low flow rates, where the reaction effects predominate and O$_2$ bubbles create a strong turbulence with occurrence of often negative counter-flows, it was not possible to estimate Q$_{out}$ for each set Q$_{in}$ and to perform least square fits (data not shown). Under these conditions no theories can be elaborated due to the chaotic flow generated.

**3.3 Mixing experiments**

Based on the results of the above model, we performed mixing experiments under hydrodynamic pumping regime at Q$_{in}$ = 50 µL min$_{-1}$ to favor conditions with dominance of the transport mechanism of the capsules by the background fluid. Because of low Reynolds number in microfluidic channels, involving laminar and uniaxial fluid flow, the mixing between streams in the flow is purely diffusive. Hence, short resident time (R$_t$) and low mixing efficiency can result when the interface between fluid streams lies at the center of a T-shape microchannel [56]. As shown in Figure 5a, in the absence of any turbulence source, two aqueous solutions (H$_2$O and eosin containing 5M H$_2$O$_2$), both injected at 50 µL min$_{-1}$ flow rate, clearly show no mixing effect in the optical images acquired in correspondence of the inlet region (R$_{t1}$ = 0 s), where the solutions are clearly separated. Unlike the T-junction region, a good mixing was observed in the middle of the T-shaped serpentine microreactor (corresponding to R$_{t3}$ = 3.24 s, see Figure 5e). This is also confirmed by the intensity profile plotted in Figure 5d, black curve. Indeed, a sharp transition of the color profile from 0 % to 100 % (in the half-section channel indicated by the arrow in Figure 5a) is straightforward at R$_{t1}$, while the transition is more gradual at R$_{t3}$ (Figure 5e and black plot of Figure 5h). To evaluate the mixing efficiency, the coefficient of variation (CoV) was measured at R$_{t1}$ and at R$_{t3}$. It was found a larger value (i.e., CoV= (0.77 ± 0.04)) as expected in the former case and, at R$_{t3}$, CoV decreased to 0.50 ± 0.15. For the calculation we considered several positions in the regions of interests (described by the boxes in Figure 5i) for three devices as described in the experimental section. To check if the microreactor can homogeneously mix the streams in presence of turbulence phenomena, pristine PMCs or catalytic PMCs were introduced into the microchannels. The intensity profiles were measured as for the reference sample at $R_{t1}$ (Figures 5b and c) where the two flows meet for the first time, and at $R_{t3}$ where the mixing effect appears good from the optical images (Figures 5f and g).

The red solution, containing 5M $H_2O_2$ and B3 (Figure 5b) or the most reactive batch B1 (Figure 5c), was injected at $Q_{in}$ = 50 µL min$^{-1}$ and mixed along the serpentine microreactor with $H_2O$ injected from the other inlet at the same speed. By comparison with the same experiment in absence of PMCs (Figure 5a), it seems that the addition of PMCs immediately shifts the fluids interface far from the middle of the channel at the T-junction level. In both cases (for B3 and B1), we also observed that the transition of color concentration (red and blue curves, Figure 5d) was less sharp with respect to the pure aqueous solution (black curve, Figure 5d) at $R_{t1}$. In these cases, the CoV values were 0.74 ± 0.12 and 0.52 ± 0.13 for B3 and B1, respectively. The mixing efficiency was further improved at $R_{t3}$. Indeed, the fluids interfaces were dissipated in the optical images (Figure 5f and g) and the color transitions lost the sharp step typical of the reference sample (see blue and red plots in Figure 5h). This was also confirmed by the decrease of CoV to 0.11 ± 0.02 for B3 and to 0.11 ± 0.01 for B1. Then, the viscoelastic effects due to the PMCs caused a perturbation of the laminar regime so that a good enough mixing occurred at $R_{t3}$. In the case of reactive PMCs, large error bars can be observed in the concentration profile (blue curves of Figure 5d, h and n), which are indicative of the large flow perturbation resulting from the catalytic reaction.

It's noteworthy to say that, when the reactive batch B1 flowed, fluids mixing was interestingly complete at the first channel meander ($R_{t2}$ = 0.5 s, Figure 5m), thanks to a combined effect of the chaotic elastic flow and massive propulsion phenomena. In this region, a substantial difference was observed in the optical images and in the concentration plot of B1 (Figures 5m and blue curve in Figure 5n) with respect to B3 (Figure 5l and red curve in Figure 5n). Indeed, beyond showing large error bars, B1 also exhibited a not steep profile at $R_{t2}$ and its CoV decreased to 0.14 ± 0.03 with respect to B3 (0.40 ± 0.09).

These experiments clearly demonstrate that PMCs enhance mass transport in fluid streams with a positive effect on the mixing efficiency in laminar regime. Compared with the intensity profile and the coefficient of variation of pure $H_2O$, less steep color concentration curves were observed and smaller CoVs were calculated at the same residence time. Similar CoVs (ranging between 0.4-0.5) were observed for the reference sample, B3 and B1 at $R_{t1}$, $R_{t2}$ and $R_{t3}$ respectively. The mixing in presence of the catalytic reaction is already efficient at $R_{t2}$ with a CoV = 0.14±0.03 comparable to that one of B3 at $R_{t3}$ (i.e., CoV = 0.11±0.02). This fact is extremely positive, since it experimentally demonstrates that micromotors are useful for improving diffusion-dominated processes in confined environments with practical applications in LOCs and OOCs.

**Conclusions**

The work aims at demonstrating the effect of viscoelastic soft micromotors on diffusion processes in microfluidic environments.

As model system, pristine PMCs and catalytic PMCs (PMCs embedding a polyoxometalate into a shell structure) were studied.

The micromotors presented in this study use a Ru-substituted polyoxometalate ($Ru_4POM$) that act as catalytic trigger to dismutate hydrogen peroxide into water and oxygen. Notably, hydrogen peroxide is considered toxic to the in vivo environment and unsuitable for use in biological applications, such as LOCs and OOCs [57, 58]. Furthermore, oxygen at high concentration could interfere with cell metabolism in organ-on-a-chip (OOC) applications. However, thanks to the high versatility of the Layer-by-Layer (LbL) assembly, when required, the $Ru_4POM$ catalyst could be easily replaced by other catalyst-fed systems (i.e. metal and photocatalytic systems for water splitting) even generating other products (i.e. acid-activated micromotors containing carbonate particles, already used for in vivo drug delivery in the gastric tract of a mouse model, or enzyme-based micromotors [32]) or micromotors responding to physical stimuli (light, acoustic or magnetic [59]). In this way, the combination of a biologically compatible catalyst with the LbL assembly method still allows for efficient mixing while avoiding altering the cells survival and metabolism.

Indeed, to understand the influence of viscoelastic micromotors on the fluid motion, PMCs were first allowed to flow, in hydrodynamic pumping conditions, across a set-up consisting of microfluidic sensors connected to a microchannel. The monitoring of some key physical parameters (pressure drop and volumetric flow rate) allowed us to shade light on the mechanisms that perturb laminar fluid regime in presence of PMCs. The effect of viscoelasticity and of self-propulsion at different injection flow speeds have been correlated to capsule-induced segmentation of continuous flow and flow recirculation. The behavior of non-reactive PMCs B3 was found in agreement with other viscoelastic systems. On the other side, activated PMCs (B1 and B2) showed an unpredictable fluid behavior as generators of turbulence phenomena due to their self-motion and oxygen bubbles production.

Then, mixing experiments demonstrated, as a positive consequence of the viscoelastic properties of the PMCs and of the chaotic flow, an increased efficiency of mixing inside a microreactor, with a significant reduction of the coefficient of variation CoV at $R_{t3}$ (0.11 for both B1 and B3), compared to pure water (0.50). Furthermore, the use of catalytic PMCs (B1) allowed an efficient mixing (CoV of 0.14) already at $R_{t2}$.

On the basis of these results, we believe that viscoelastic micromotors can be considered an interesting tool to improve mass transport in experimental conditions purely dominated by diffusion. In particular, they could be useful to enhance mixing between fluids containing reagents or nutrients in LOC and OOC platforms to achieve a rapid mixing, a fast diagnosis or to guarantee efficient transport of nutrients for cell survival.

**Acknowledgements**

Published in Chemical Engineering Journal (20 November 2019, 123572). DOI: https://doi.org/10.1016/j.cej.2019.123572

The authors thank the "Cluster in Bioimaging" (cod. QZYCUM0, "Aiuti a sostegno dei cluster tecnologici regionali 2014", Bando Regione Puglia n. 399 del 28/07/2014) and FISR project – C.N.R. 'Tecnopolo di Nanotecnologia e Fotonica per la Medicina di Precisione' – CUP B83B17000010001 Tecnomed. The authors are grateful to Mrs E. Perrone of CNR-NANOTEC for technical support.

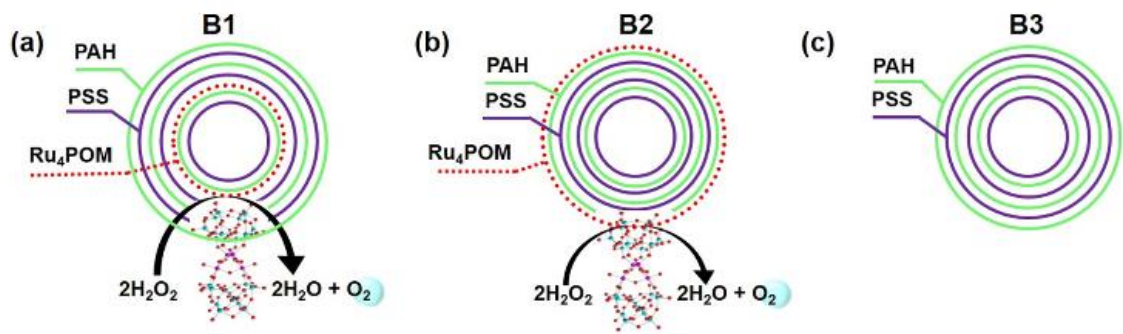

**Fig. 1.** Reactive batches with Ru4POM adsorbed in polyelectrolyte multilayers as (a) inner layer (B1), (b) external layer (B2); (c) non-reactive batch of polyelectrolyte multilayers (B3). The oxygen was produced from the dismutation reaction of B1 and B2 with H2O2. For sake of clarity only six out of ten total polyelectrolyte layers are represented in the shells of B1, B2 and B3.

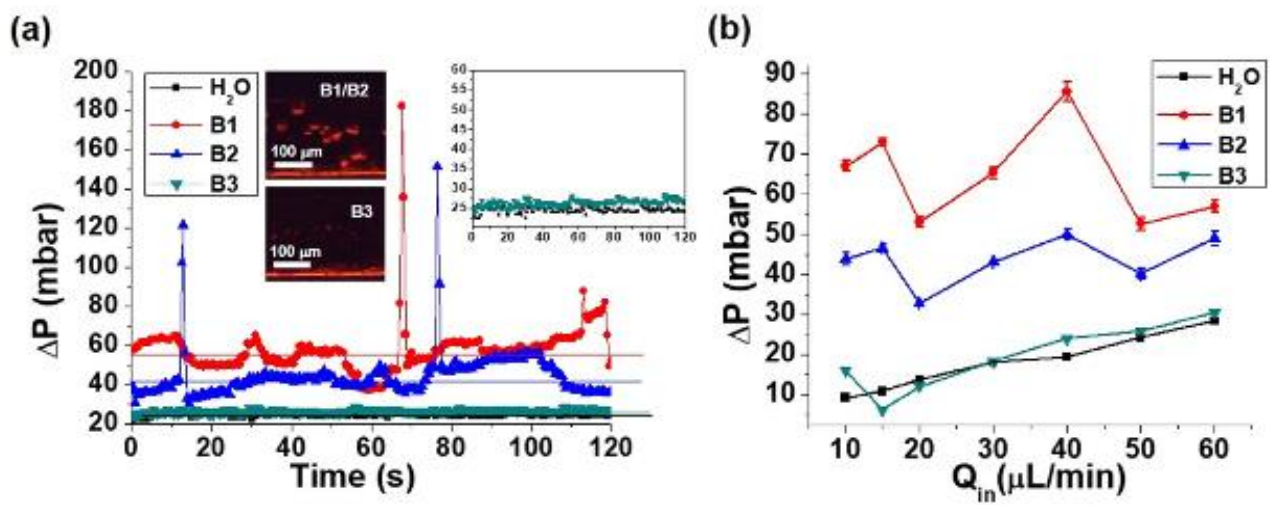

**Fig. 2.** (a) Representative curves of the temporal evolution of the pressure drop (ΔP): $Q_{in}$ = 50 μL/min for B1 (red symbols), B2 (blue symbols) and B3 (cyan symbols), pure water (black symbols). The sample consists of an aqueous solution containing $1,9*10^3$ cps per μL and 5 M $H_2O_2$. Linear fit is also drawn for each set of data. In the left inset, fluores- cence images of B1/B2 and B3 mixed with $H_2O_2$ and flowing in a microchannel: $O_2$ bubbles are clearly visible in the first case. In the right inset, zoomed view of ΔP for B3 and $H_2O$: slight differences/fluc- tuations for increasing flow rate are visible in the nearly overlapping. (b) ΔP values versus $Q_{in}$ of mi- crocapsules batches (B1, B2, B3) evaluated as de- scribed in the text. The curves associated to water are also shown as reference. (For interpretation of the references to color in this figure legend, the reader is referred to the web version of this article.)

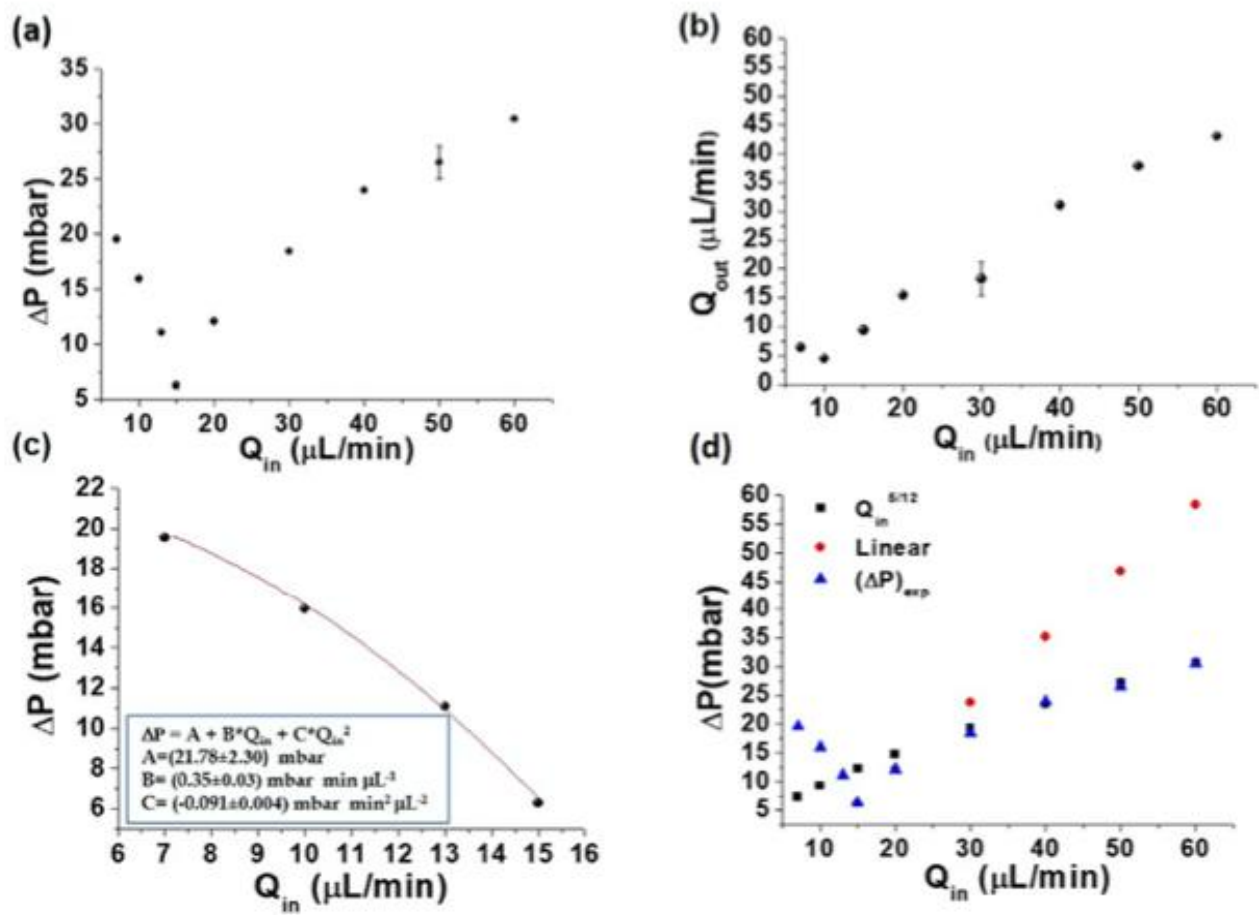

**Fig. 3.** B3 hydrodynamic characterization. (a) Measured pressure drop (ΔP) and (b) outlet volumetric flow rate ($Q_{out}$) vs inlet volumetric flow rate ($Q_{in}$). (c) Measured ΔP, at $Q_{in} \leq 15$ μL/min, following a quadratic polynomial law (fitting is reported as red continuous line). (d) ΔP profile decomposed in different functional forms depending on the $Q_{in}$ range: linear trend for $15 < Q_{in} \leq 25$ μL/min (red symbols), power (5/12) dependence for $Q_{in} > 25$ μL/min (black symbols). (For interpreta- tion of the references to color in this figure legend, the reader is referred to the web version of this ar- ticle.).

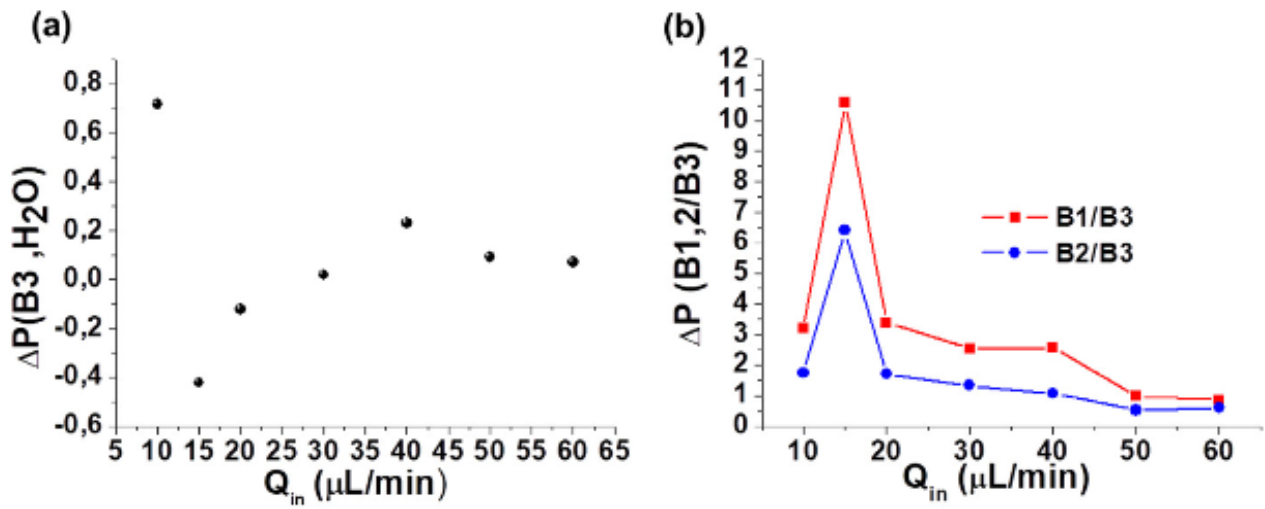

**Fig. 4.** (a) Pressure drop of B3 normalized to the one of the transport fluid (ΔP(B3,H$_2$O)) according to formula (1). (b) Pressure drop of B1 and B2 normalized to the one of B3 (ΔP(B1,2/B3)) (formula (2)).

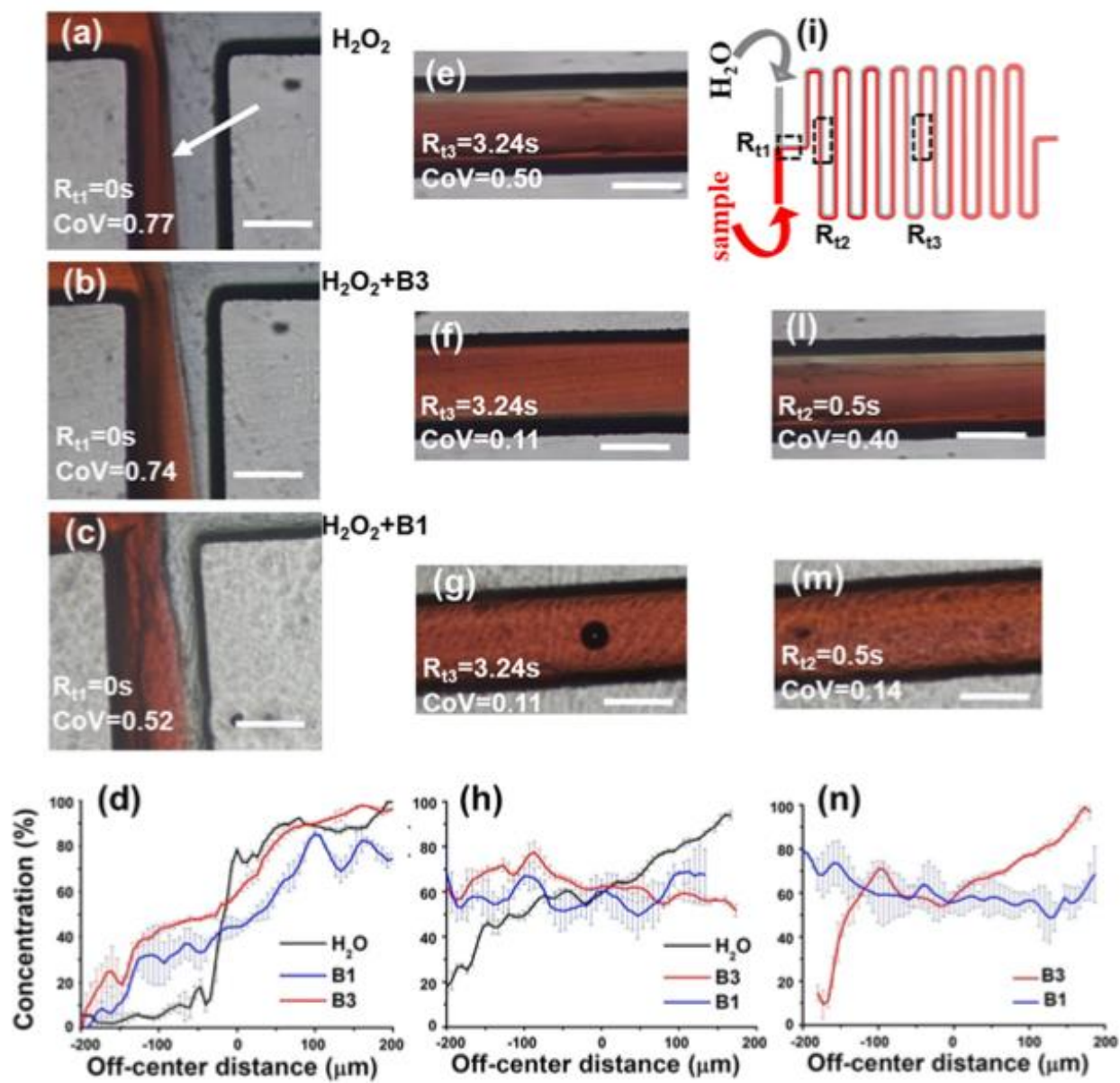

**Fig. 5.** Mixing tests in a serpentine microreactor using pure H2O and a colored aqueous solution injected at $Q_{in}$ = 50 μL/min. The mixing efficiency (CoV) is reported at different residence times ($R_{t1}$, $R_{t2}$, $R_{t3}$) corresponding to the dashed black squared location along the serpentine (i). The scale bar in the optical images is 400 μm. In absence of PMCs the two streams show: (a) a laminar flow (CoV = 0.77) at $R_{t1}$ = 0 s, with a sharp transition in the color concentration plot (black curve, d), and (e) a mixing effect (CoV = 0.50) at $R_{t3}$ = 3.24 s, with a gradual intensity transition (black curve, h). Tests with B3 dissolved in the red solution exhibited: (b) a shift of the fluid-fluid interface far from the middle of the channel with a CoV = 0.74 at $R_{t1}$ and (f) a good mixing (CoV = 0.11) at $R_{t3}$. Tests

with B1 also showed: (c) a shift of the fluid-fluid interface with CoV = 0.52 at $R_{t1}$ and (g) an efficient mixing (CoV = 0.11) at $R_{t3}$. For B3 and B1 a less steep color concentration (red and blue curves respectively in (d) and (h)) can be observed both at $R_{t1}$ and $R_{t3}$. (l, m, n) The mixing in presence of the catalytic reaction is already efficient at $R_{t2}$ = 0.5 s (CoV = 0.40 for B3 and CoV = 0.14 for B1). (For interpretation of the references to color in this figure legend, the reader is referred to the web version of this article.)